\documentclass[a4paper, 10pt, conference]{IEEEtran}

\IEEEoverridecommandlockouts                   

\usepackage{graphicx}
\usepackage{color}
\usepackage{amsmath}
\usepackage{multirow}
\usepackage{booktabs}
\usepackage{url}
\usepackage{amsmath}
\usepackage[position=t,singlelinecheck=off]{subfig}
\usepackage{adjustbox}
\usepackage{subfloat}
\usepackage[english]{babel}
\usepackage{tikz}
\usetikzlibrary{positioning}

\newcommand\copyrighttext{%
  \footnotesize \textcopyright 2012 IEEE. Personal use of this material is permitted. Permission from IEEE must be obtained for all other uses, in any current or future media, including reprinting/republishing this material for advertising or promotional purposes, creating new collective works, for resale or redistribution to servers or lists, or reuse of any copyrighted component of this work in other works. This paper has been accepted for publication in the 2020 Twelfth International Conference on Quality of Multimedia Experience (QoMEX).}
\newcommand\copyrightnotice{%
\begin{tikzpicture}[remember picture,overlay]
\node[anchor=south,yshift=10pt] at (current page.south) {\fbox{\parbox{\dimexpr\textwidth-\fboxsep-\fboxrule\relax}{\copyrighttext}}};
\end{tikzpicture}%
}

\begin{document}

\title{User Experience of Reading in Virtual Reality - Finding Values for Text Distance, Size and Contrast}

\author{
 \IEEEauthorblockN{Tanja Koji\'c$^1$, Danish Ali$^1$, Robert Greinacher$^1$, Sebastian M\"oller$^{1,2}$, Jan-Niklas Voigt-Antons$^{1,2}$}
 \IEEEauthorblockA{$^1$Quality and Usability Lab, TU Berlin, Germany\\ $^2$German Research Center for Artificial Intelligence (DFKI), Berlin, Germany}
}

%\IEEEpubid{\makebox[\columnwidth]{978-1-7281-5965-2-0/20/\$31.00 \copyright 2020 IEEE \hfill} \hspace{\columnsep}\makebox[\columnwidth]{ }}

\maketitle
\copyrightnotice

%%%%%%%%%%%%%%%%%%%%%%%%%%%%%%%%%%%%%%%%%%%%%%%%%%%%%%%%%%%%%%%%%%%%%%%%%%%%%%%%
\begin{abstract}
Virtual Reality (VR) has an increasing impact on the market in many fields, from education and medicine to engineering and entertainment, by creating different applications that replicate or in the case of augmentation enhance real-life scenarios. Intending to present realistic environments, VR applications are including text that we are surrounded by every day. However, text can only add value to the virtual environment if it is designed and created in such a way that users can comfortably read it. 
With the aim to explore what values for text parameters users find comfortable while reading in virtual reality, a study was conducted allowing participants to manipulate  text parameters such as font size, distance, and contrast.  
Therefore two different standalone virtual reality devices were used, Oculus Go and Quest, together with three different text samples: Short (2 words), medium (21 words), and long (51 words). 
Participants had the task of setting text parameters to the best and worst possible value. Additionally, participants were asked to rate their experience of reading in virtual reality. 
Results report mean values for angular size (the combination of distance and font size) and color contrast depending on the different device used as well as the varying text length, for both tasks. Significant differences were found for values of angular size, depending on the length of the displayed text. However, different device types had no significant influence on text parameters but on the experiences reported using the self-assessment manikin (SAM) scale.

\end{abstract}

% Keywords
\begin{keywords}
    Virtual Reality, Readability, User Experience, Text Parameters 
    \vspace{-2.5em}
\end{keywords}

%\begin{tikzpicture}[overlay, remember picture]
%\path (current page.north) node (anchor) {};
%\node [below=of anchor] {%
%2020 Twelfth International Conference on Quality of Multimedia Experience (QoMEX)};
%\end{tikzpicture}

%%%%%%%%%%%%%%%%%%%%%%%%%%%%%%%%%%%%%%%%%%%%%%%%%%%%%%%%%%%%%%%%%%%%%%%%%%%%%%%%
\section{INTRODUCTION}

The number of Virtual reality (VR) applications is rapidly growing, and nowadays, virtual environments are used in various fields such as medicine, engineering, education, design, training, and entertainment \cite{onyesolu2011understanding,martin2017virtual}. In all those sectors, everyday work entails reading a lot of text \cite{karim2007reading}, and by replicating parts of this work environments into virtual reality should as well include it. There are many use cases where text could be used inside of virtual reality. To create better realism while replicating the real world or only to create additional value on top where it would not be possible in real life. However, reading in virtual reality is not yet focus of activities in industry and research as much as it is creating realistic models and environments. Still, it is hard to imagine a real-world without text, and poor readability in virtual reality might also lead to the bad overall quality of experience, similarly as it is the case for web or mobile interfaces \cite{hall2004impact}. 
When it comes to defining good or bad experience of readability in virtual reality, there is to our knowledge little research done so far. 
Guidelines for how best to display text are not yet established and are based on recommendations published by companies such as Google or Unity, but standardized recommendations are not yet available.

\subsection{Related work}

Readability as a term is used within different domains so that it can be referred to as the accuracy of reading \cite{ardoin2005accuracy}; as the ability to understand the text with taking into account the speed of reading \cite{pikulski2002readability}; or to the visual representation of the characters themselves \cite{ali2013reading}. In each of these applications, the reader must be able to read and understand the text.
One of the essential prerequisites for a good user experience of reading text from a screen is that visual information is designed and displayed clearly and comfortably \cite{roufs1997text}. That property of allowing sentences to be read from a given material, regardless of their meaning, is crucial for readability. Efficient readability requires excellent legibility of the displayed text, where legibility refers to the visual properties, meaning how easy it is to recognize a character or a symbol \cite{zuffi2007human}. That is why concerning readability in virtual reality, the first step is to focus on legibility and visual representation of text. 

Visual representation of text is equally important in digital solutions, as it is in printed forms, and a lot of standardization and research was done in this area when it comes to desktop website \cite{vu2006web} or mobile \cite{bandeira2010towards} content. One of the most prominent standards, when it comes to ensuring the accessibility of website contents, is WCAG2.0  \cite{yates2005web}. The idea behind this standardization is to ensure that users can understand, browse and interact with websites. Among other suggestions that are proposed by WCAG2.0, there is an indication for contrast ratio between text color and background color depending on the text size. 
WCAG2.0 Level A means that the website information is accessible to all users and that the conforming alternate version is available. Level AA requires a contrast ratio of at least 4.5:1 for standard text and 3:1 for large text. Large text is defined as minimally 18 points or 14 points bold. Level AAA requires a contrast ratio of at least 7:1 for standard text and 4.5:1 for large text.  

One of the characteristics of the user interface (UI) in virtual reality is that the screen size is not limited \cite{bastug2017toward}. Therefore, VR brings the advantage when it comes to space for representing information in all directions. However, apart from ability to present more information, it is important to define the position of elements around the user so that the user can comfortably comprehend information. In order to achieve that, three zones are defined for the user interface in VR: \textit {content zone} is comfortable 140$^{\circ}$-wide angle in front of the user; the \textit{peripheral zone} is left and right to the user between 70$^{\circ}$ and 105$^{\circ}$ angle; and everything else that is behind the user is named as \textit{curiosity zone} \cite{alger2015visual}. Also, three zones are defined for the distance between user and interface, where information should not be placed closer than 0.5m and beyond 20m.

As for displaying text in virtual reality, not only the font size of characters is important, but also the distance in which this text is displayed. Google has proposed a new font size unit named distance independent millimeter - dmm. It is defined as height of the character of 1mm on the distance of 1m. That was proposed as angular size had no standardized unit so far in virtual reality, and it is giving a consistent view on any screen at any distance. Therefore, the unit was quickly adopted and is used by both industries and in research. 
One of the research studies that is reporting in dmm is a study that explored basic parameters to measure text box size, text size, and specific preferences \cite{dingler2018vr}. The results are reporting that the angular size of the text should be 41dmm +/- 14 dmm. Additionally, participants reported preference of having white text on black background over black text on a white background and displaying text in sans-serif type font (Arial) over serif (Times New Roman). However, in this study, only Oculus Rift as a head-mounted display (HMD) was connected to a personal computer and providing high resolution, but not different resolutions or standalone HMDs were considered. 
Another study that was conducted regarding readability in VR explored performing daily routine work in VR. Instead of text applications, they created static images and asked the user to read the text and interact, like with a web browser or emails. Additionally, the performance was measured by reading speed and accuracy \cite{grout2015reading}. Results show that distance to imaginary representation is important in VR and that permanent head movement helps to improve readability. However, in this study, participants had no possibility of adjusting font parameters by themselves. 

\subsection{Objectives} 
In order to explore what are the text parameters for comfortable reading and displaying text in virtual reality, a study was designed, allowing participants to set up their preferred settings. The focus of the study was on three main parameters: the size of text, the distance between user and text, and the color contrast of text and background. Although research has already been done in the area of readability for web \cite{caldwell2008web}, there had been little focus on how those standards apply for text inside of virtual reality, especially for standalone VR devices. Differentiation of conditions for a study has been created by using different text lengths (short tittle, medium and longer paragraph), as well as different HMD devices with different screen resolutions. The remaining sections of the present paper explain the process of creating the application for reading and study, as well as the test set up with the aim to answer these questions:
\begin{itemize}
    \item What are the text parameters that users choose as the best and the worst for reading in virtual reality? 
    \item Are selected text parameters significantly different depending on text length? 
    \item Are selected text parameters significantly different depending on virtual reality HMD devices?
    \item What influence do device type and text length have on reported user experience while reading in virtual reality? 
\end{itemize}

\section{METHODS}

\subsection{Test setup}
The study was conducted in a separate university's lab room equipped with two different head-mounted displays (HMD) - Oculus Go and Oculus Quest, as they have different specifications \cite{hillmann2019comparing}. The resolution of Oculus Go is 1280x1440 LCD, while Oculus Quest has a resolution of 1440x1600 OLED. Difference between devices is also in degrees of freedom, and for Oculus Go it is only 3, while Oculus Quest has 6 degrees of freedom. As a virtual environment, the Unity application had been created with a simplistic empty environment, where text could be in focus. The virtual environment had three different stages of adjusting text distance, font size and contrast of text and background. In order not to influence participants with any other text in the virtual environment, as guidance though stages just colored objects and sounds were used. Further on, there were different types of text displaying depending on the number of words: short (2 words), medium (21 words) and long (51 words). The font used for displaying text was Arial, with regular weight and default spacing. 
In application, it was possible to manipulate with text settings in order to set distance of text (between 0 and 10000mm), font size (from 5pt to 40pt) and contrast ratio (from 1:1 to 21:1). All manipulations were done by using controllers connected with head-mounted displays, and the same gestures were used in each stage to simplify gameplay for participants who are not experienced with virtual reality. Even though in each stage participant could set up only one value, it was possible to switch and move between stages in order to create final results as wanted. 
With the aim to find out what font settings are preferred, participants had to set the best and the worst possible combination of distance, size and contrast of each text on both devices. Once when participants selected settings, all values were saved in a log file of the application. The order of text and devices was balanced across participants. 

\begin{figure}[h!]
\centering
\includegraphics[width=0.5\textwidth]{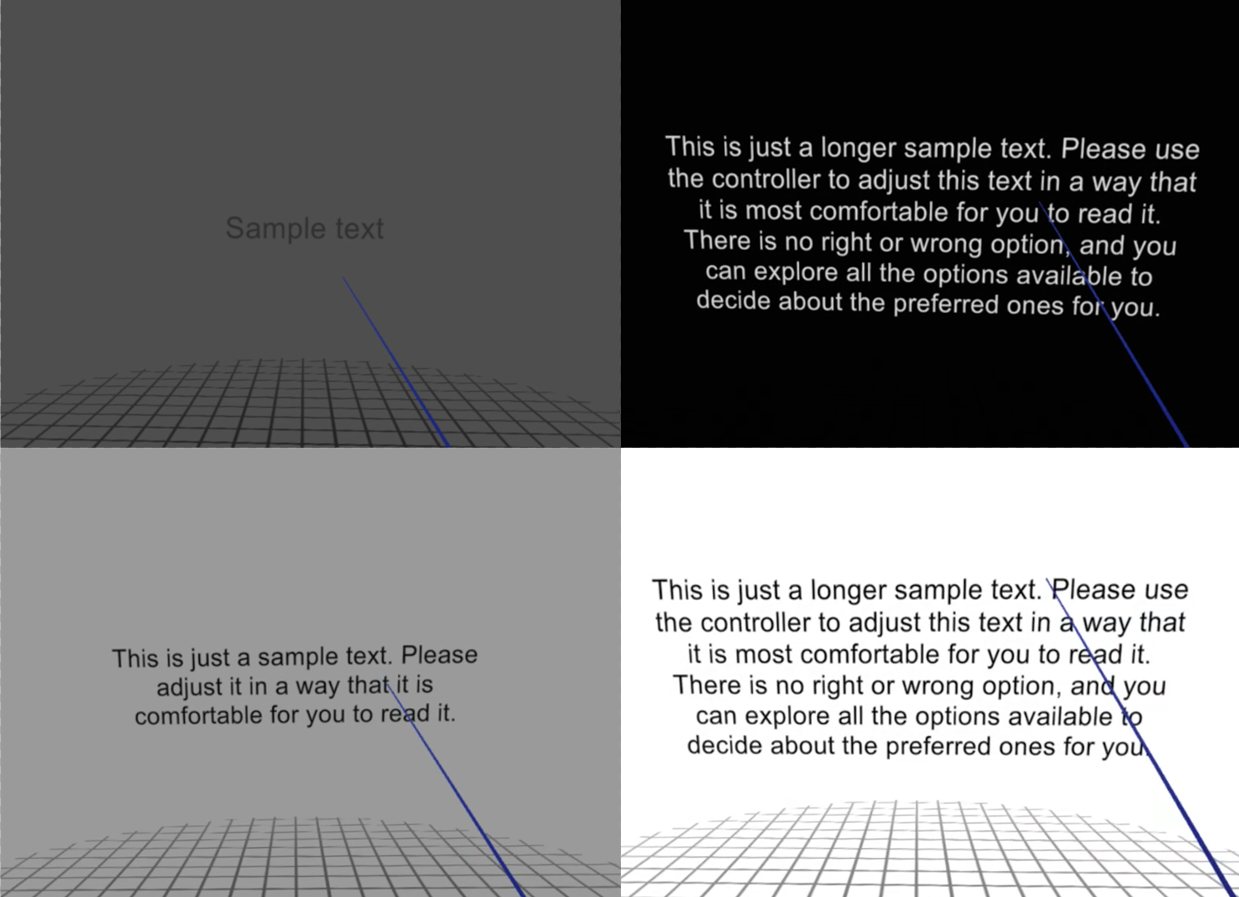}
\caption{Examples of user interface inside of VR application where user could use controllers to manipulate text parameters (text size, distance and color contrast).}
\label{fig:text}
\vspace{0em}
\end{figure}

\begin{figure}[h!]
\centering
\includegraphics[width=0.5\textwidth]{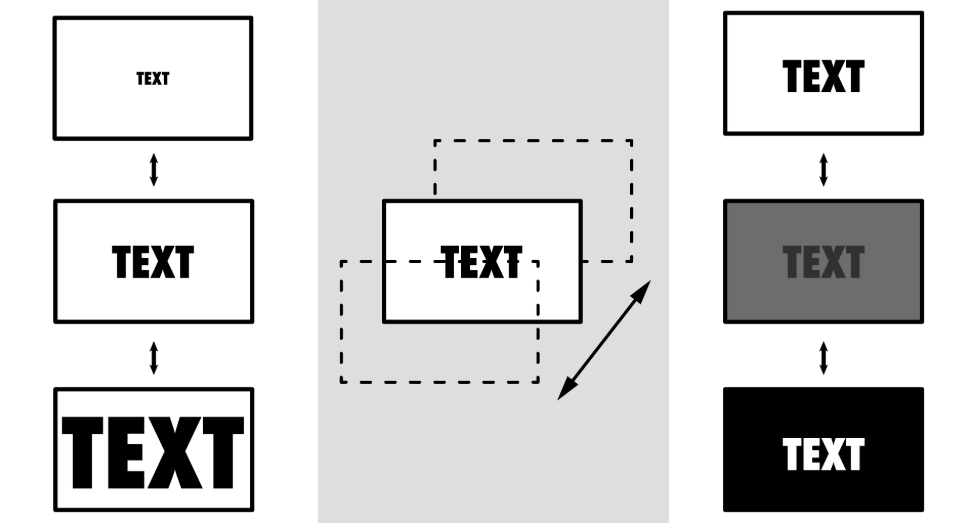}
\caption{Graphical explanation of possibilities of manipulations with text parameters: text size, distance and contrast between text and background}
\label{fig:text}
\vspace{-2.5em}
\end{figure}

\subsection{Procedure}
Participants were invited to the lab room at university, each at different separated time slot in order to take part in the experiment alone. At the beginning, the participant was welcomed by a moderator and presented with an introduction of the study. After signing a consent, a participant was given a pre-questionnaire asking about demographic information, their previous experience with virtual reality, and daily habits in reading as well as if they have any kind of eyesight improvements such as glasses or lenses. Additionally, participants were asked to fill in the Affinity for Technology Interaction (ATI) Scale questionnaire \cite{franke2019personal} in order to find out their tendency to actively engage in intensive technology interaction.
Next step was that participant was given in the introduction to each of Oculus devices, controllers and interactions needed for the experiment. As part of the introduction, each participant tried to use both devices with a training application that had the same interactions. However, no data was saved so that the participant could explore all possibilities and options of application and interactions with controllers. Also, at this stage, participants could change settings of HMD that fit them with the best comfort. Participants with glasses could also wear their classes below the head-mounted display as Oculus is providing enough space for it. 

After training was done, the participant had to set up text parameters for each condition. Altogether, there were six conditions, as all combinations of two Oculus devices (Go and Quest) and three different lengths of text (2, 21 and 51 words). For each condition, a participant had to set up text parameters - distance, font size and contrast of text.
A participant was asked to do two tasks: one was to set up text parameters that readability of the text is the best possible, and the second was to set up parameters of text to the worst possible readability. Those tasks were randomized, as well as conditions of the study. After each task, a participant was filling in the Self-Assessment Manikin (SAM) \cite{bradley1994measuring} as an emotion assessment tool that uses graphic scales, depicting cartoon characters expressing three emotion elements: pleasure, arousal and dominance. Finally, after settings text parameters for all conditions, the participant was asked to rate the usability of reading in virtual reality with the System Usability Scale (SUS) \cite{brooke1996sus}. The final questionnaire was asking participants to evaluate general readability in virtual reality when they were using their best text settings and to leave any additional comments. 

\subsection{Participants}
A complete set of 22 participants (54.5\% male and 45.5\% female) was participating in the study. The average age of the participants was 28.41 years (SD = 9.56 , min = 19, max = 62). The majority of participants had some experience with virtual environments (86.4\%), while only 3 participants (16.6\%) never tried virtual reality before, and nobody was an expert with virtual reality. Additionally, participants reported that on average, the majority reads between 2 and 4 hours a day (77.3\%). Less than 1 hour of a day reads 13.6\% of participants, while only 9.1\% reads between 5 and 7 hours a day. Lastly, participants reported if they have any eyesight improvements while participating in a study. Eye lenses were worn 3 participants (13.6\%), 6 participants (27.3\%) were using glasses, and 13 other participants (59.1\%) did not need eyesight improvement. 

\section{RESULTS} 

A repeated measure Analysis of Variance (ANOVA) was performed to determine statistically significant differences. An overview of all significant effects that will be explained in the following sections is given with table \ref{tab:results}. Additionally, table \ref{tab:mean} is showing for all conditions mean values that participants have selected as the best settings of text parameters of size, distance and color contrast.  

\begin{table}[htbp]
  \centering
  \caption{Mean values of angular size and contrast ration over all conditions for the best settings of text parameters. Difference in conditions was by changing length of text - 2 words (short), 21 words (medium), 51 words (long) and device (Oculus Go and Quest). }
  \resizebox{\columnwidth}{!}{
    \begin{tabular}{lllcl}
    \toprule
\multicolumn{1}{l}{Parameter} & \multicolumn{1}{l}{Text Lenght} & \multicolumn{1}{l}{Device} & \multicolumn{1}{c}{Mean} & \multicolumn{1}{l}{SD}
    \\
    \midrule
    Angular size & Short & Go & $27.48$     & $9.86$   \\
    Angular size & Short & Quest & $32.06$     & $18.75$   \\
    Angular size & Medium & Go & $17.17$     & $4.90$   \\
    Angular size & Medium & Quest & $16.18$     & $5.25$   \\
    Angular size & Long & Go & $16.01$     & $9.25$   \\
    Angular size & Long & Quest & $17.87$     & $24.52$   \\
    Contrast ratio & Short & Go & $11.89$     & $6.68$   \\
    Contrast ratio & Short & Quest & $10.12$     & $7.32$   \\
    Contrast ratio & Medium & Go & $10.47$     & $7.52$   \\
    Contrast ratio & Medium & Quest & $8.71$     & $5.94$   \\
    Contrast ratio & Long & Go & $11.02$     & $7.73$   \\
    Contrast ratio & Long & Quest & $9.37$     & $5.75$   \\
    \bottomrule
    \end{tabular}%
    }
  \label{tab:mean}%
\end{table}%

\begin{table}[htbp]
  \centering
  \caption{Effects of different HMD devices (Device) and text length (Length) on angular size (dmm) as combination of font size and distance and SAM dimension of dominance (SAM\_D), in different tasks where users had to select the best text settings (Pos) and the worst (Neg).}
  \resizebox{\columnwidth}{!}{
    \begin{tabular}{lllcllrl}
    \toprule
\multicolumn{1}{l}{Effect} & \multicolumn{1}{l}{Parameter} & \multicolumn{1}{l}{Cond} & \multicolumn{1}{c}{$df_{\textnormal n}$} & \multicolumn{1}{l}{$df_{\textnormal d}$} & \multicolumn{1}{c}{$F$} & \multicolumn{1}{c}{$p$} & \multicolumn{1}{c}{$\eta_{\textnormal G}^2$}
    \\
    \midrule
    dmm & Lenght & Pos & $1$     & $22$    & $9.89$ & $ .005$ & $0.32$ \\
    SAM\_D & Lenght+Device & Neg & $1$     & $22$    & $6.14$ & $ .022$ & $0.22$ \\
    \bottomrule
    \end{tabular}%
    }
  \label{tab:results}%
\end{table}%

\subsection{Angular size}
\begin{figure}[h!]
\centering
\includegraphics[width=0.5\textwidth]{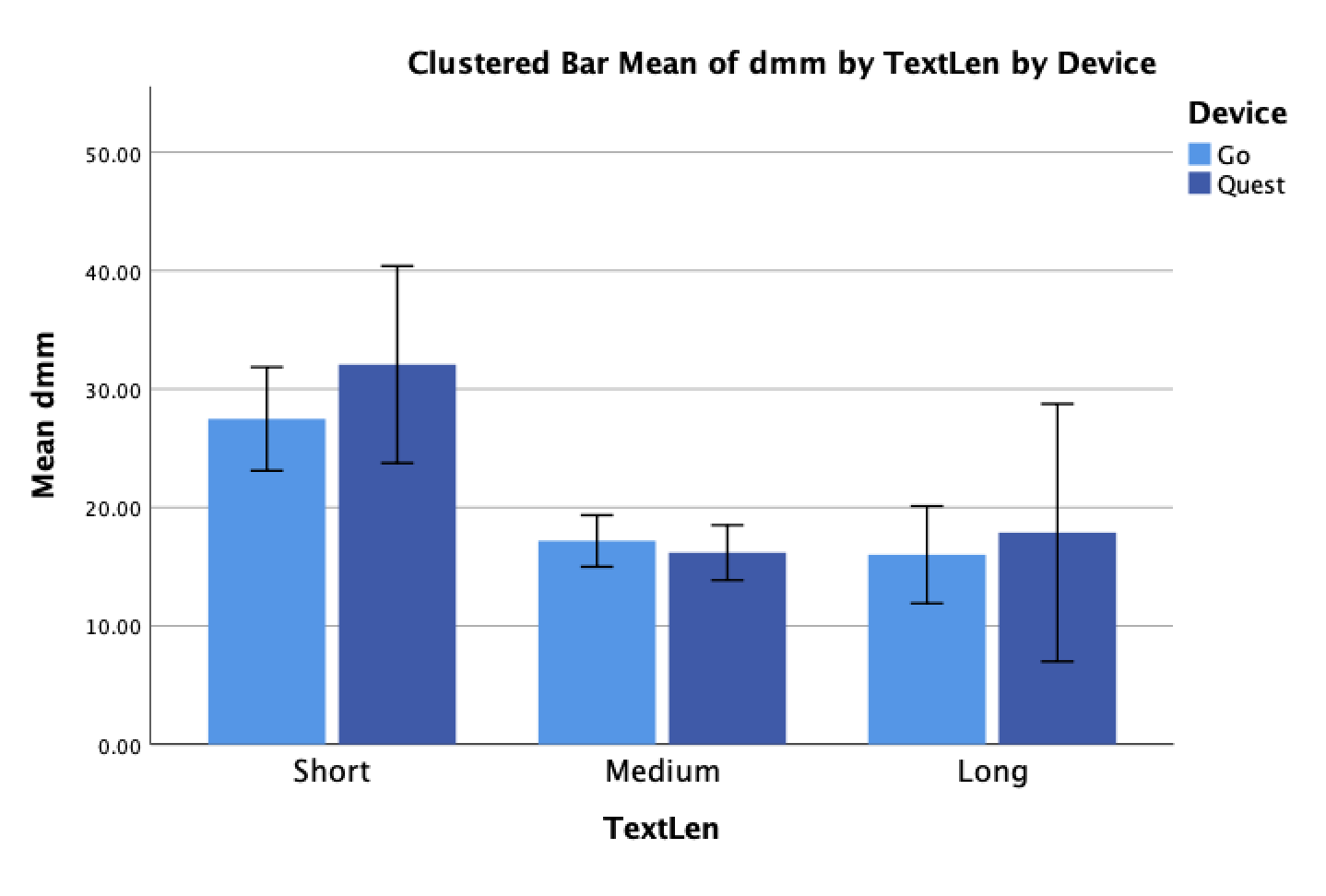}
\caption{Mean value of angular size (in dmm) for task of choosing the best text parameters for reading in VR, in relation to length of text display and type of VR device used. }
\label{fig:dmm}
\vspace{-2em}
\end{figure}

\begin{figure}[h!]
\centering
\includegraphics[trim={0 1.2cm 2.5cm 0cm},clip,width=0.5\textwidth]{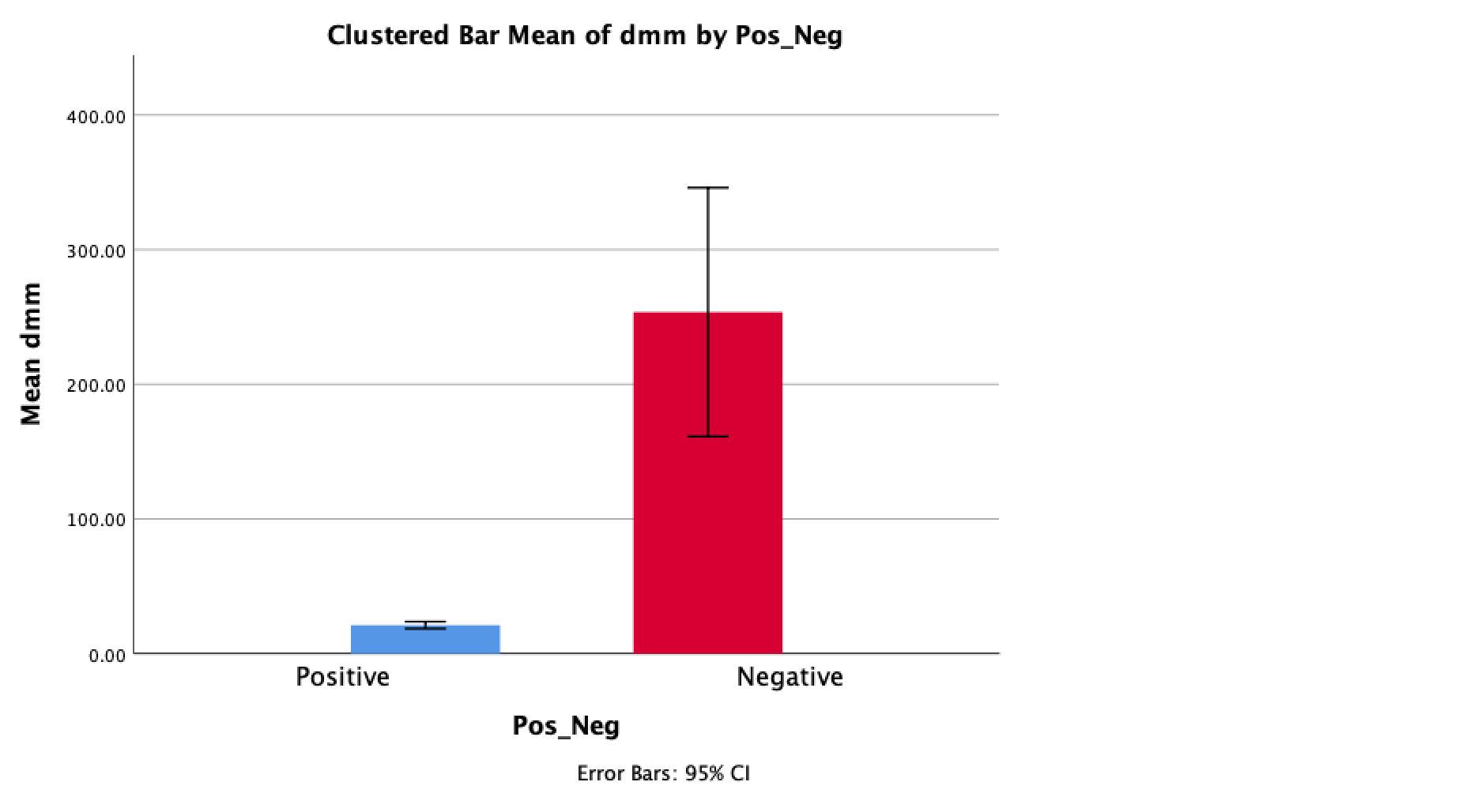}
\caption{Mean values of angular size (in dmm) for both tasks, choosing the best (positive) and the worst (negative) text parameters for reading in VR.  }
\label{fig:dmm_neg}
\vspace{-3em}
\end{figure}

Text parameters of font size and text distances have been combined and calculated to represent one value named angular size. Angular size is described using Google's unit for angular size called distance-independent millimeter (dmm), where 1dmm equals 1mm height at 1m viewing distance. The main effect found for the task where participants had to select the best text parameters is about text length significantly influencing the preferred angular size of the text. This result is represented by figure \ref{fig:dmm} and all mean values are shown in table \ref{tab:mean}. Results show that for short text length, angular size is significantly bigger compared with when the text that was displayed as medium or long. It can also be seen that this effect is valid for both devices equally, and that device type had no significant influence on angular size. 
When it comes to the task where participants had to choose settings of text parameters to create the worst experience of reading in VR, no significant influences were found with the influence of device or text length. However, values for each condition are significantly higher compared to positive task (figure \ref{fig:dmm_neg}) for both Oculus Go with short (M=297.1191 SD=691.48351), medium (M=203.5112 SD=389.74864) and long text (M=203.5112 SD=389.74864); as well as for Oculus Quest with short (M=404.5926 SD=866.13453), medium (M=253.1436 SD=365.15349) and long (M=169.5190 SD=235.94466) text.

\subsection{Contrast ratio}

\begin{figure}[h!]
\centering
\includegraphics[width=0.5\textwidth]{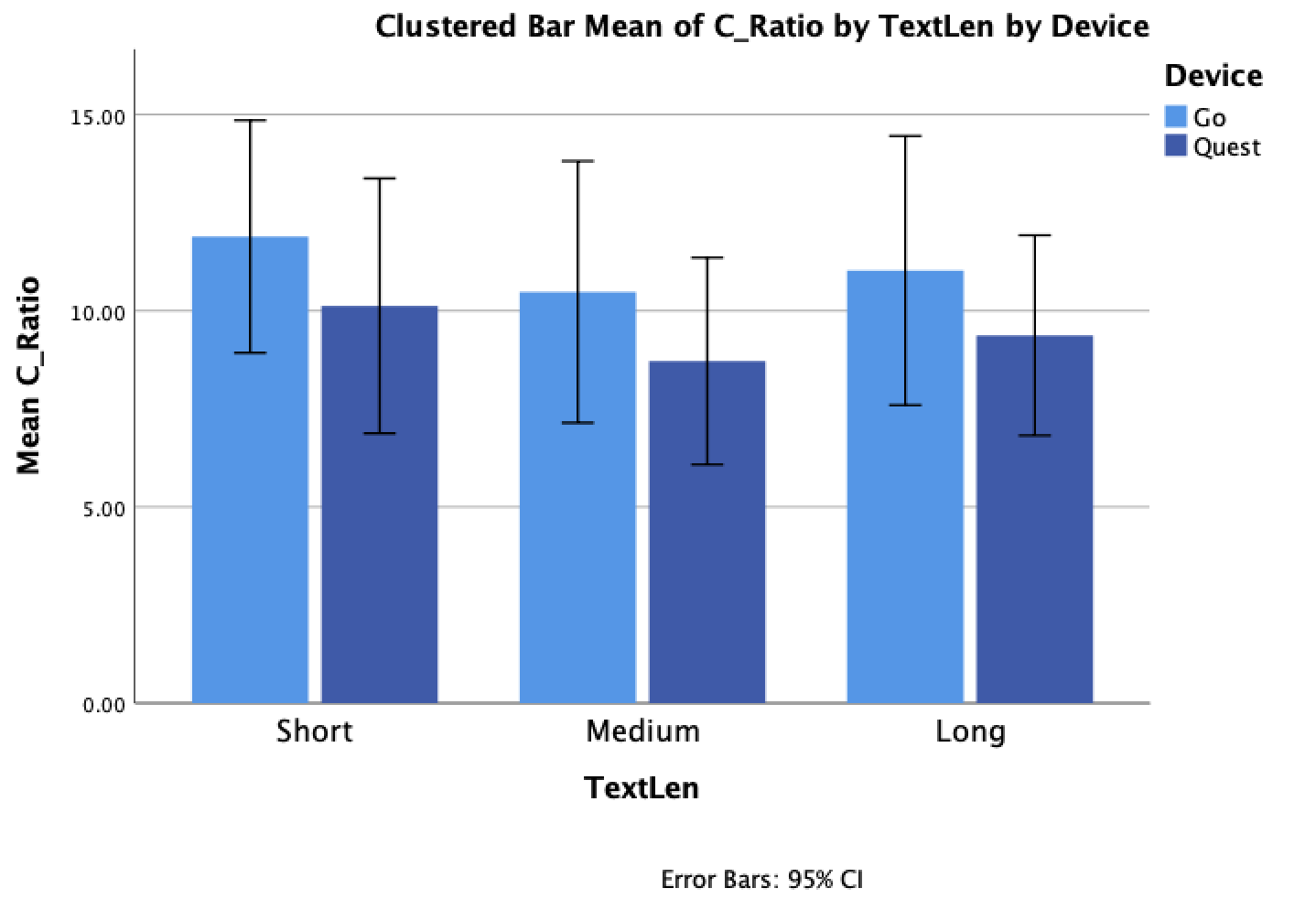}
\caption{Mean value of contrast ratio between color of text and background for task of choosing the best text parameters for reading in VR, in relation to length of text display and type of VR device used. }
\label{fig:contrast}
\vspace{+1em}
\end{figure}

\begin{figure}[h!]
\centering
\includegraphics[width=0.5\textwidth]{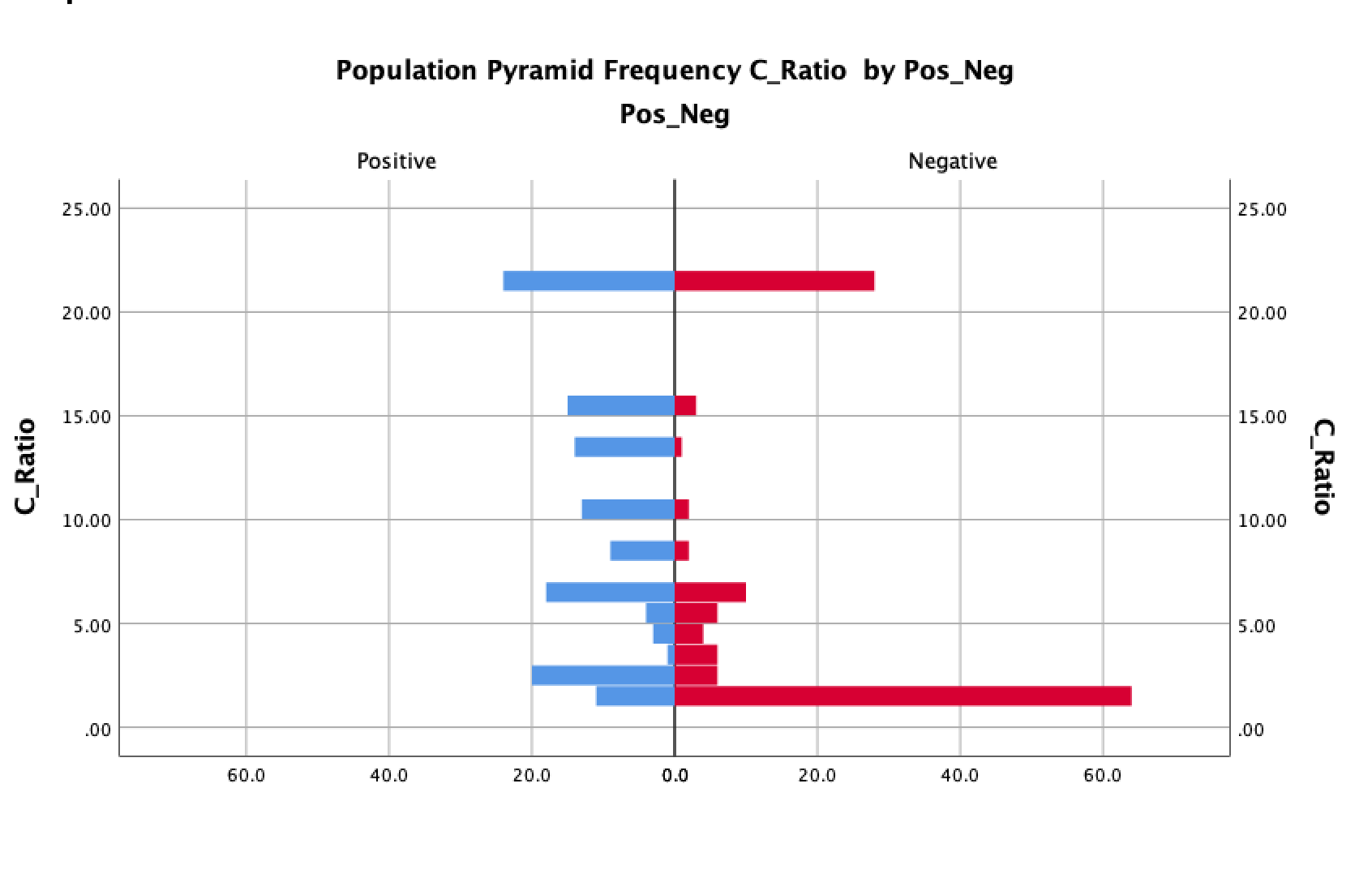}
\caption{Frequency of choosing values of contrast ratio between color of text and background for both tasks, choosing the best (positive) and the worst (negative) text parameters for reading in VR.}
\label{fig:c_ratio_neg}
\vspace{0em}
\end{figure}

Contrast ratio was calculated between the color of text and the color of the background of a virtual environment, and all mean values are reported in table \ref{tab:mean} for settings that participants have selected as preferred ones for reading in VR and shown with figure \ref{fig:contrast}. Even though text length and device type had no significant influence, mean values when it comes to the negative example of reading in VR are interestingly different for both devices. Oculus Go showing short (M=6.65, SD=8.24), medium (M=6.8677, SD=8.05033) and long text (M=8.59, SD=9.19) or Oculus Quest and short (M=7.48, SD=7.85), medium (M=6.45, SD=7.06) and long (M=6.04, SD=7.02) text. A comparison of contrast ratio and how frequently it was selected in each task is shown with figure \ref{fig:c_ratio_neg}.

\subsection{SAM}

\begin{figure}[h!]
\centering
\includegraphics[width=0.5\textwidth]{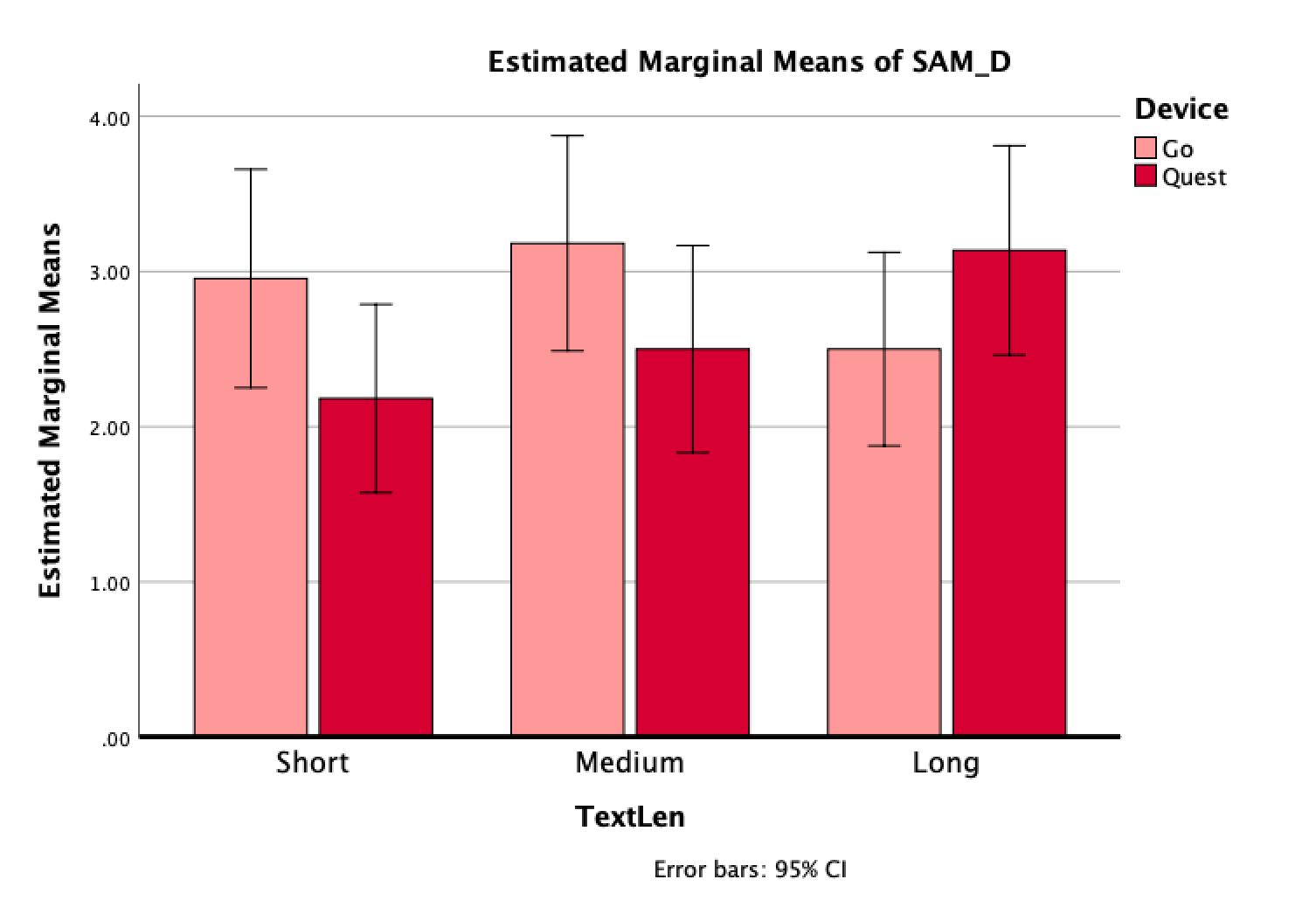}
\caption{Mean value of SAM dominance dimension for the task of choosing the worst text parameters considering reading in VR, in relation to the type of VR device used. }
\label{fig:sam_d}
\vspace{-1em}
\end{figure}

For the task where participants had to set up parameters to the worst representation of text, it can be observed in figure \ref{fig:sam_d}  that length of the text and device type influenced how users have rated SAM dimension dominance. While seeing bad examples for longer text, participants have rated that feel significantly more in control while using device Oculus Quest (M=3.13, SD=1.52) compared to the same long text while using device Oculus Go (M=2.50, SD=1.40). However, rated feeling of dominance was significantly higher with device Oculus Go for short (M=2.95, SD=1.58) and medium text (M=3.18, SD=1.56) compared to using Oculus Quest with same short (M=2.18, SD=1.36) and medium (M=2.50, SD=1.50) text. 

%missing SUS value!!! 

\section{DISCUSSION}
This study intended to explore comfortable reading settings of text parameters for users, with different lengths of text and using different devices. Therefore participants were asked to select the best possible setting in order to create the most comfortable user experience of reading for them. Additionally, the task was also to set text parameters to create a negative experience of reading by choosing the worst parameters. Accordingly, the results of the study are showing mean values of text size, distance (combined as angular size) and color contrast between text and background that users prefer. 

The difference in mean values of angular size between two tasks is as assumed significantly different. However, it is interesting to notice how in the negative condition mean values were in general much higher \ref{fig:dmm_neg}. Users were asked to set the worst combination of text size and distance, and were not anyhow influenced by what to choose - the option of very small or big text. However, as results show, many participants have decided on creating the option with very big angular size with choosing text font to be the biggest possible at a very close distance. Even though standards such as WCAG2.0 \cite{yates2005web} usually suggest font size for normal text to be larger than a specific value, it might be important to find and define as well the maximum value. %Value for the angular size of text in virtual reality to which users feel that text is readable and their user experience is good. 

When it comes to the task of setting the best possible settings for angular size, results show that the length of text is significantly influencing preferred font size and distance. A short text that was only 2 words in this study, and might have been comprehended by some participants as a title resulted in significantly larger angular size compared to values for medium (21 words) and long (51 words) text. Even though it was not told to participants to think of as title or paragraph, mean values of short text for both Oculus Go (M=27.48dmm) and Oculus Quest (M=32.06dmm) are close by values to the ones that Google guidelines are suggesting for headlines (40dmm) or titles (32dmm). Likewise, medium or long text mean values are just a little bit below the proposed Google guidelines values for body or caption text.  

Further on, contrast ration was calculated for each condition between colors of text and background. Ratio 1:1 represents a situation where the same color of the text is represented on the background of that color and therefore looks invisible. On the other side ratio 21:1 is representing the highest possible contrast, for example, black text on a white background or white text on black background. The mean results of contrast ratio for the task where participants had to select the best possible text settings are not significantly different for different text lengths or devices, but in all conditions are at least 7:1, that requires WCAG2.0 Level AAA. 
However, even though the WCAG2.0 standard is suggesting guidelines by providing a ratio that has to be minimally fulfilled, interesting results were reported about the ratio that is closer to maximal values. 

For the task where participants had to set the text parameters to create a negative reading experience, the maximal contrast ratio of 21 was selected as the worst by around 1/3 of cases. Therefore creating the design guidelines with only the minimum value required for contrast ratio might not always be the right solution for representations of text in virtual reality. Especially while designing an environment where  background is very bright, as a screen of virtual reality is so close to the eye that bright color might feel like too much light in the eye and overall experience would be perceived as rather a negative one. However, results from this study are just indicating this trend but are not yet able to define the values at which this effect would start happening. Reading only the mean values for the contrast ratio of the task with setting text parameters to create a negative reading experience is not enough. Moreover, it is important to notice that the arrangement of data is rather on very low or very high contrast ratio value. 

Finally, when it comes to user experience of reading in virtual reality and how people have perceived it, both together device type and text length had significant influence on results. Participants have reported that the feeling of being in control, measured by SAM dimension of dominance, is higher for longer text while using Oculus Quest compared to the Oculus Go where participant reporter higher feeling on control for shorter text when text was set by the worst parameters. It might be due to the different resolutions of screens between devices, as well as different degrees of freedom that devices are providing. Interestingly, devices had an influence when it comes to reading with the negatively set text parameters, even though no significant difference was found when choosing the best possible text settings.

\section{CONCLUSION}

This paper is presenting the values selected by participants as the best settings of text parameters of size, distance and color contrast, as well as the values for the worst settings concerning reading text in virtual reality on standalone headsets. 
With the aim to investigate the relation between user experience and reading in virtual reality, this paper is presenting values for basic text parameters - angular size and color contrast ratio. Results are presenting user preferences as well as what users defined as a negative experience of reading in virtual reality. With a task to select the best and the worst possible settings of text parameters, there are significant differences between those two, as well as the influence of text length and different standalone HMD devices. 
In summary, this paper is presenting which experience, positive or negative, can be expected when setting basic text parameters to specific values.

It is interesting to observe that so far guidelines have been proposing values for contrast ratio and text size as minimal required. 
It is not only the minimum value that is important, but the guidelines should also consider proposing maximal values to which users feel comfortable when it comes to angular size and contrast ratio for reading in virtual reality. Angular size was rater significantly higher by participants in task with setting the worst parameters, compared to the task with choosing the best font size and distance for reading in virtual reality. Even though virtual reality is providing bigger space to represent content, when it comes to text, having bigger font, just because the display and virtual environment are allowing it, was not preferred by users. Similarly, the contrast ratio that had maximal values was also not preferred by users and was rated by some users as the worst set of text parameters. However, when considering the contrast ratio, it is also important to address the question of what color is being used in background and how bright is the virtual environment. Those two questions are not addressed by results in this paper and are an exciting aspect that can be researched in future work. 

Study task about choosing the best set of text parameters resulted in values that can be used as guidelines for angular size and contrast ratio. Results can be used for designing a reading experience in virtual reality, especially while designing an experience for standalone HMD device. Even though no significant difference was found when it comes to the device types, it might be taken into consideration that the difference between them could have been bigger. However, when it comes to the length of text, results have shown that the angular size for short text, such as title, is set by participants as significantly higher compared to the angular size for medium or long text. 

All the reported values are showing the set up of the best of the worst settings of text parameters (size, distance and contrast) for reading in virtual reality but are not reporting on minimal and maximal values that would be acceptable without destroying the user experience. Therefore, this would be needed future work, which could lead to better defining not only minimal requirements for angular size and contrast ration, but also their maximal limits. Further on, more text parameters could be included, such as different font families, different canvas types, or different spacing between letters and words. 

%Grammarly checked 

%\section{Acknowledgment}
% Example of comment and hidden Acknowledgment area.

\bibliographystyle{IEEEtran}
\bibliography{references.bib} 

\begin{thebibliography}{10}
\providecommand{\url}[1]{#1}
\csname url@samestyle\endcsname
\providecommand{\newblock}{\relax}
\providecommand{\bibinfo}[2]{#2}
\providecommand{\BIBentrySTDinterwordspacing}{\spaceskip=0pt\relax}
\providecommand{\BIBentryALTinterwordstretchfactor}{4}
\providecommand{\BIBentryALTinterwordspacing}{\spaceskip=\fontdimen2\font plus
\BIBentryALTinterwordstretchfactor\fontdimen3\font minus
  \fontdimen4\font\relax}
\providecommand{\BIBforeignlanguage}[2]{{%
\expandafter\ifx\csname l@#1\endcsname\relax
\typeout{** WARNING: IEEEtran.bst: No hyphenation pattern has been}%
\typeout{** loaded for the language `#1'. Using the pattern for}%
\typeout{** the default language instead.}%
\else
\language=\csname l@#1\endcsname
\fi
#2}}
\providecommand{\BIBdecl}{\relax}
\BIBdecl

\bibitem{onyesolu2011understanding}
M.~O. Onyesolu and F.~U. Eze, ``Understanding virtual reality technology:
  advances and applications,'' \emph{Adv. Comput. Sci. Eng}, pp. 53--70, 2011.

\bibitem{martin2017virtual}
J.~Mart{\'\i}n-Guti{\'e}rrez, C.~E. Mora, B.~A{\~n}orbe-D{\'\i}az, and
  A.~Gonz{\'a}lez-Marrero, ``Virtual technologies trends in education,''
  \emph{EURASIA Journal of Mathematics Science and Technology Education},
  vol.~13, no.~2, pp. 469--486, 2017.

\bibitem{karim2007reading}
N.~S.~A. Karim and A.~Hasan, ``Reading habits and attitude in the digital
  age,'' \emph{The Electronic Library}, 2007.

\bibitem{hall2004impact}
R.~H. Hall and P.~Hanna, ``The impact of web page text-background colour
  combinations on readability, retention, aesthetics and behavioural
  intention,'' \emph{Behaviour \& information technology}, vol.~23, no.~3, pp.
  183--195, 2004.

\bibitem{ardoin2005accuracy}
S.~P. Ardoin, S.~M. Suldo, J.~Witt, S.~Aldrich, and E.~McDonald, ``Accuracy of
  readability estimates' predictions of cbm performance.'' \emph{School
  Psychology Quarterly}, vol.~20, no.~1, p.~1, 2005.

\bibitem{pikulski2002readability}
J.~J. Pikulski, ``Readability,'' \emph{Retrieved September}, vol.~20, p. 2014,
  2002.

\bibitem{ali2013reading}
A.~Z.~M. Ali, R.~Wahid, K.~Samsudin, and M.~Z. Idris, ``Reading on the computer
  screen: Does font type have effects on web text readability?.''
  \emph{International Education Studies}, vol.~6, no.~3, pp. 26--35, 2013.

\bibitem{roufs1997text}
J.~A. Roufs and M.~C. Boschman, ``Text quality metrics for visual display
  units:: I. methodological aspects,'' \emph{Displays}, vol.~18, no.~1, pp.
  37--43, 1997.

\bibitem{zuffi2007human}
S.~Zuffi, C.~Brambilla, G.~Beretta, and P.~Scala, ``Human computer interaction:
  Legibility and contrast,'' in \emph{14th International Conference on Image
  Analysis and Processing (ICIAP 2007)}.\hskip 1em plus 0.5em minus 0.4em\relax
  IEEE, 2007, pp. 241--246.

\bibitem{vu2006web}
K.-P.~L. Vu, R.~W. Proctor, and F.~P. Garcia, ``Web site design and
  evaluation,'' \emph{Handbook of human factors and ergonomics}, pp.
  1323--1353, 2006.

\bibitem{bandeira2010towards}
R.~Bandeira, R.~Lopes, and L.~Carri{\c{c}}o, ``Towards mobile web accessibility
  evaluation,'' \emph{Free and Open Source Software for Accessible Mainstream
  Applications (FOSS-AMA), colocated with ETAPS}, pp. 27--28, 2010.

\bibitem{yates2005web}
R.~Yates, ``Web site accessibility and usability: towards more functional sites
  for all,'' \emph{Campus-wide Information systems}, 2005.

\bibitem{bastug2017toward}
E.~Bastug, M.~Bennis, M.~M{\'e}dard, and M.~Debbah, ``Toward interconnected
  virtual reality: Opportunities, challenges, and enablers,'' \emph{IEEE
  Communications Magazine}, vol.~55, no.~6, pp. 110--117, 2017.

\bibitem{alger2015visual}
M.~Alger, ``Visual design methods for virtual reality,'' \emph{Ravensbourne.
  http://aperturesciencellc. com/vr/VisualDesignMethodsforVR\_MikeAlger. pdf},
  2015.

\bibitem{dingler2018vr}
T.~Dingler, K.~Kunze, and B.~Outram, ``Vr reading uis: Assessing text
  parameters for reading in vr,'' in \emph{Extended Abstracts of the 2018 CHI
  Conference on Human Factors in Computing Systems}.\hskip 1em plus 0.5em minus
  0.4em\relax ACM, 2018, p. LBW094.

\bibitem{grout2015reading}
C.~Grout, W.~Rogers, M.~Apperley, and S.~Jones, ``Reading text in an immersive
  head-mounted display: An investigation into displaying desktop interfaces in
  a 3d virtual environment,'' in \emph{Proceedings of the 15th New Zealand
  Conference on Human-Computer Interaction}.\hskip 1em plus 0.5em minus
  0.4em\relax ACM, 2015, pp. 9--16.

\bibitem{caldwell2008web}
B.~Caldwell, M.~Cooper, L.~G. Reid, and G.~Vanderheiden, ``Web content
  accessibility guidelines (wcag) 2.0,'' \emph{WWW Consortium (W3C)}, 2008.

\bibitem{hillmann2019comparing}
C.~Hillmann, ``Comparing the gear vr, oculus go, and oculus quest,'' in
  \emph{Unreal for Mobile and Standalone VR}.\hskip 1em plus 0.5em minus
  0.4em\relax Springer, 2019, pp. 141--167.

\bibitem{franke2019personal}
T.~Franke, C.~Attig, and D.~Wessel, ``A personal resource for technology
  interaction: development and validation of the affinity for technology
  interaction (ati) scale,'' \emph{International Journal of Human--Computer
  Interaction}, vol.~35, no.~6, pp. 456--467, 2019.

\bibitem{bradley1994measuring}
M.~M. Bradley and P.~J. Lang, ``Measuring emotion: the self-assessment manikin
  and the semantic differential,'' \emph{Journal of behavior therapy and
  experimental psychiatry}, vol.~25, no.~1, pp. 49--59, 1994.

\bibitem{brooke1996sus}
J.~Brooke \emph{et~al.}, ``Sus-a quick and dirty usability scale,''
  \emph{Usability evaluation in industry}, vol. 189, no. 194, pp. 4--7, 1996.

\end{thebibliography}

%\addtolength{\textheight}{-3cm}

\end{document}